\documentclass[sigconf]{acmart}
\AtBeginDocument{%
  }

\setcopyright{acmlicensed}
\copyrightyear{2026}
\acmYear{2026}

\acmConference[JAWs '26]{Journal Ahead Workshop}{August,
  2026}{Munich, Germany}
\usepackage{array}

\newif\ifcomments
\commentsfalse    

\ifcomments
    \newcommand{\yuan}[1]{\textcolor{purple}{\textit{[Yuan: #1]}}}
    \newcommand{\tim}[1]{\textcolor{blue}{\textit{[Tim: #1]}}}
    \newcommand{\mariam}[1]{\textcolor{brown}{\textit{[Mariam: #1]}}}
\else
    \newcommand{\yuan}[1]{}
    \newcommand{\tim}[1]{}
    \newcommand{\mariam}[1]{}
\fi

\begin{document}

\title{BurnRiSc: Toward Non-Invasive Burnout Screening in Open Source from Public Repository Signals}


\author{Timofey Sanko}
\affiliation{%
 \institution{Queen's University}
 \city{Kingston}
 \state{Ontario}
 \country{Canada}}
\email{timofey.sanko@queensu.ca}

\author{Yuan Tian}
\affiliation{%
  \institution{Queen's University}
  \city{Kingston}
  \state{Ontario}
  \country{Canada}
}
\email{y.tian@queensu.ca}

\author{Mariam Guizani}
\affiliation{%
  \institution{Queen's University}
  \city{Kingston}
  \state{Ontario}
  \country{Canada}
}
\email{mariam.guizani@queensu.ca}

\renewcommand{\shortauthors}{Sanko et al.}

\begin{abstract}
Burnout is a chronic occupational syndrome, and open source is close to a worst case for it: maintainers absorb unbounded demand with no manager to reallocate work and no organization to notice decline. The cost is not only personal. Burnout precedes withdrawal, and in projects sustained by a handful of maintainers, one departure can break infrastructure that thousands of downstream systems depend on. Yet the field has no way to see it coming: self-report inventories, the only existing measure, miss exactly the contributors most in need of detection and cannot be applied retroactively, so the field cannot even ask how common burnout is or what helps. 

We present BurnRiSc, a framework that operationalizes the Oldenburg Burnout Inventory's two dimensions, exhaustion and disengagement, as 14 behavioral and linguistic signals computed from GitHub activity and scored against each contributor's own history. The signals aggregate into two weighted dimension scores, with weights learned from labeled cases, and average into a monthly Burnout Risk Score (BRS). In a preliminary evaluation across 68 contributors in ten repositories (ten disclosed burnout cases, twelve comparable-volume collapses, and 46 comparison contributors), sustained BRS elevation precedes 6 of 10 disclosures by 6-15 months, 8 of 10 when adding peak BRS as a second criterion, and 10 of 10 over any prior time frame. We thus present BurnRiSc as evidence that burnout is screenable from public data.

\end{abstract}




\maketitle

\section{Introduction}\label{sec:introduction}

Burnout is a chronic occupational syndrome, recognized by the WHO as arising from unmanaged workplace stress~\cite{who2019burnout}. Job Demands--Resources theory attributes it to sustained demands unmatched by available resources ~\cite{demerouti2001job}. Open source exhibits this configuration structurally: maintainers absorb demand that arrives without limit from users who bear no cost in generating it, with no manager who can reallocate work, no organization positioned to notice a decline, and no compensation that would justify absorbing it indefinitely. Community surveys report burnout as widespread and largely unaddressed~\cite{raman2020stress,godliauskas2025well,heath2025burnout}.

The cost is not only personal. Burnout is associated with anxiety, depression, and physical fatigue ~\cite{maslach2001job}, and among software professionals it predicts attrition~\cite{moore2000one,trinkenreich2024predicting}. In open source, this compounds: many projects depend on very few core contributors~\cite{avelino2016novel}, and their departure is a leading route to project abandonment~\cite{coelho2017modern,avelino2019abandonment}. Maintainer burnout is therefore a software sustainability problem as much as a well-being one.

Despite these stakes, empirical work on burnout in OSS remains thin, largely because of a lack of measurement. Burnout is inferred through validated instruments such as the Maslach Burnout Inventory~\cite{maslach1997maslach} and the freely available Oldenburg Burnout Inventory (OLBI)~\cite{demerouti2001job}, which scores two dimensions, exhaustion and disengagement, from self-report items. In OSS, these instruments face three main limitations. First, disengagement is half of what OLBI measures, so contributors furthest into burnout, and those who already left, are the least likely to respond. Second, an inventory measures state at the moment of administration, so cases that have already occurred are permanently invisible to it. The third concerns what can be studied at all. Because burnout has no repository-computable form, it cannot enter the mining-software-repositories toolkit that empirical SE otherwise applies to contributor behavior. CHAOSS, the Linux Foundation's open source health-metrics project, defines a Project Burnout metric but specifies it entirely through surveys and interviews, in contrast to the trace-computable metrics that make up most of its catalogue~\cite{chaossburnout}. Every study must instead recruit participants directly, which bounds samples at the scale of recruitment rather than of available data and forecloses the questions a measurable construct would open: whether maintainers' burnout risk rises after a popularity surge or a CVE, whether it falls when a project joins a foundation or secures sustained funding, and whether interventions such as adding maintainers or adopting AI coding agents measurably reduce it.

Software engineering has not ignored contributor well-being so much as measured what it could reach directly: affect. Mäntylä et al. show that valence, arousal, and dominance extracted from contributor communication track productivity~\cite{mantyla2016mining}, and related work studies sentiment in code review~\cite{ahmed2017senticr} and toxicity in OSS interactions~\cite{miller2022did,ferreira2021shut}. Affect captures how a contributor feels in a given exchange. However, burnout is an occupational syndrome accumulated over months of demands exceeding resources, and a contributor can be exhausted and disengaged while writing perfectly civil comments. \textbf{We therefore ask: can a validated burnout inventory be approximated from traces contributors already leave in public?}

To answer this question, we present BurnRiSc, which operationalizes OLBI's exhaustion and disengagement dimensions as 14 behavioral and linguistic signals computable from GitHub activity: timestamps, throughput, and text drawn from commits, pull requests, and issues. Each signal is scored relative to a contributor's own history rather than against an absolute threshold, since contribution volume in OSS varies by orders of magnitude across roles. The weights combining signals within each dimension are learned from labeled cases via the L-BFGS-B optimization algorithm~\cite{Byrd1995} rather than assigned by the authors, so the mapping from signal to dimension answers to evidence rather than intuition. The two dimension scores are then averaged into a monthly Burnout Risk Score (BRS) for each target contributor.

We evaluate BurnRiSc on 68 contributors across ten repositories: ten who publicly disclosed burnout, twelve whose contribution volume collapsed comparably, and 46 comparison contributors. Three findings emerge. Sustained BRS elevation precedes 6 of the 10 disclosures within a 24-month window; adding peak BRS as a second criterion raises coverage to 8 of 10, and every disclosure is preceded by at least one period of elevated BRS outside that window. Weight fitting from equal-weight, four anchor starts shows that off-hours activity and commit volume are dominant signals. Peak BRS correlates only weakly with an independent activity measure ($r = 0.26$), so the score is not a proxy for contribution activity. 

The contributions of this paper are as follows:

\begin{itemize}
\item We argue that OLBI's dimensions are approximable from passively observable repository signals, and give a 14-signal operationalization grounded in psychological and linguistic theory.

\item We develop BurnRiSc, a scoring pipeline that indexes each contributor against their own history and learns dimension weights from labeled cases to form its monthly Burnout Risk Score (BRS).

\item We evaluate BurnRiSc on 68 contributors, including ten self-disclosed cases, showing that BRS rises months ahead of disclosure and separates cases from comparison contributors.

\item We discuss the use case for BurnRiSc, its implications, and our plan to expand on this work. The code for implementing BurnRiSc and a supplementary document are available at: \url{https://github.com/RISElabQueens/BurnRiSc}.
\end{itemize}

We submit this work to JAWs for early feedback on the framework and evaluation. We hope to incorporate the feedback and submit an extended version to TSE or TOSEM. 


\section{Background and Related Work}\label{sec:background}
\textbf{Burnout as a Construct:}
The WHO recognizes burnout as a work-related syndrome resulting from chronic, unmanaged workplace stress, not an individual pathology~\cite{ICD11, Day25}. It comprises three dimensions: emotional exhaustion, cynicism, and reduced personal accomplishment~\cite{maslach1997maslach}. Its consequences extend from the individual (anxiety, depression, physical fatigue, and social withdrawal~\cite{Gorgievski08}), to the organization (lowered morale, reduced efficiency, and decreased job satisfaction~\cite{Taris2006}). Unlike acute stress, it develops gradually and resists quick recovery~\cite{Leblond24}, which is what makes early detection valuable.

\textbf{Burnout Instruments:} Burnout is inferred through standardized instruments rather than directly observed. The Maslach Burnout Inventory (MBI) is the gold standard~\cite{maslach1997maslach, Nunarnan2026}, but it is proprietary, restricting its use at scale~\cite{Pate2023}. Additionally, its third dimension, personal accomplishment, is debated as non-essential to the syndrome~\cite{Schaufeli2005}. The Oldenburg Burnout Inventory (OLBI) addresses both concerns: it is freely available and reduces burnout to two dimensions, exhaustion and disengagement, averaged into a single score~\cite{demerouti2001job}.  
We adopt OLBI as the theoretical foundation for this framework, operationalizing each dimension through observable behavioral and linguistic signals. 
\mariam{changed to OSS since the paragraph is about OSS. Unless you would like to add some work on burnout in SE in general. We have discussed wellbeing in SE in a recent FSE paper if that helps: https://dl.acm.org/doi/abs/10.1145/3803437.3805581}
\tim{OSS is enough for this portion}

\textbf{Wellbeing in OSS:} OSS-specific work is mostly descriptive. Raman et al. found that stress and burnout are recurring themes in OSS communities~\cite{raman2020stress} and recent surveys found a lot of undetected burnout among contributors ~\cite{godliauskas2025well}. But no study measures or predicts it. Repository-based studies capture affect~\cite{mantyla2016mining, ahmed2017senticr} and toxicity in OSS interaction~\cite{Ehsani25}. This is adjacent to, but distinct from, burnout. It reflects how a contributor feels at a given time, not the chronic exhaustion and disengagement that accumulate over months of unsustainable workload. No prior work operationalizes a validated burnout inventory into passive computable signals; BurnRiSc fills that gap.
\mariam{if more space is needed we can turn the subsection in sec 2 into bold text}
\section{The BurnRiSc Framework}\label{sec:framework}
\subsection{Overview and Applicability Conditions}\label{sec:framework_condition}
BurnRiSc takes as input the public activity history of a contributor within a repository and returns a monthly Burnout Risk Score (BRS) that ranges from  0 to 1 throughout a contributor's tenure. BRS is built on five categories of timestamped activity available through the GitHub API: commits with messages, pull request creation/merge events, code reviews, pull request review comments, and issue comments. 
Two conditions must hold for a contributor to qualify for BRS computation. First, the account must belong to a human contributor, excluding bots and agentic contributors (identified by the \texttt{[bot]} tag). Second, the contributor must have sufficient history: BRS scores each signal against a contributor's past, so a minimum tenure is required for percentile ranks and trend slopes. In this study, we use a cautious threshold of three years, as the level component requires only three prior monthly observations to compute a valid percentile rank, and the trend component requires a four-month window; three years provides substantial margin above these minimums to ensure percentile ranks and trend estimates are stable rather than merely computable.

\subsection{Operationalizing OLBI: Signal Catalog}
Self-report instruments do not scale to OSS: they require recruitment and sustained participation from contributors who are distributed, pseudonymous, and under no obligation to respond. We therefore operationalize each OLBI dimension, i.e., exhaustion and disengagement, through seven proxy variables computable from public activity traces. Starting from each dimension, we identify manifestations reported in the burnout literature, then define a metric that captures each from collected data. Table~\ref{tab:catalogue} lists the resulting 14 signals. Since burnout presents differently for every individual~\cite{Marin12, Edwards98, Runcan13}, the aim is to use a variety of variables to indicate that a contributor is starting to experience burnout. While these variables in isolation do not signal burnout, each is a proxy for either exhaustion or disengagement, respectively. Proxy variables relating to exhaustion focus on markers of fatigue and possible overworking. Proxy variables relating to disengagement aim to catch withdrawal. As with the OLBI, signals are combined and averaged to produce a single score, i.e., the \textit{Burnout Risk Score (BRS)}. A threshold determined by the data (\S\ref{sec:fitting}) is used to predict burnout risk. The BRS is not a diagnosis: it signals that a contributor may benefit from being checked in on and offered support.

\begin{table*}
\setlength{\abovecaptionskip}{4pt}
\caption{ Signal catalog. Each proxy variable is mapped to an OLBI dimension,
a direction, and a theoretical basis. Direction indicates how the raw metric
moves under burnout: $\uparrow$ increase, $\downarrow$ decrease.}\tim{citations added}
\label{tab:catalogue}
\small
\begin{tabular}{@{}llcp{0.29\linewidth}p{0.26\linewidth}@{}}
\toprule
\textbf{Signal} & \textbf{Source} & \textbf{Dir.} & \textbf{Computation} & \textbf{Theoretical Basis}\\
\midrule
\multicolumn{5}{@{}c}{\textbf{\textit{Exhaustion} — emotional, physical, and cognitive fatigue \cite{demerouti2001job}} } \\
\cmidrule(l{0pt}r{0pt}){1-5}
\addlinespace[2pt]
Off-hours Activity (OA)  & Timestamps & $\uparrow$ & Proportion of activity outside inferred work window (10-hour sliding window) & Circadian disruption reduces sleep quality and increases fatigue \cite{Khan18,Harma06}.\\
\addlinespace[2pt]
Linguistic Arousal (LA)  & Texts & $\downarrow$ & Mean VAD arousal \cite{Jia25, VAD} score from the NRC-VAD lexicon& Lower arousal reflects mental fatigue  \cite{mantyla2016mining, Warriner13,  Zajenkowska2015}.\\
\addlinespace[2pt]
PR Merge Time (MT)       & PRs & $\uparrow$ & Average days from PR creation to merge & Slower throughput indicates lower performance and reflects exhaustion \cite{Behrens23}.\\
\addlinespace[2pt]
Lexical Exhaustion (LE)  & Texts & $\uparrow$ & Mean similarity to exhaustion seed phrases (\S\ref{sec:text}) & Explicit fatigue language reflects exhaustion.\\
\addlinespace[2pt]
Sentiment--Exhaustion (SE) & Texts & $\downarrow$& Distance from neutral sentiment class (fine-tuned RoBERTa, \S\ref{sec:text}) & As contributors get exhausted, their emotional expression becomes more neutral \cite{Zapf99}.\\
\addlinespace[2pt]
Lexical Diversity (LDv)  & Texts & $\downarrow$ & Moving-average type--token ratio \cite{Bestgen25}& Reduced vocabulary richness reflects cognitive exhaustion \cite{Harrison97, Killgore2010}.\\
\addlinespace[2pt]
Hapax Rate (HR)          & Texts & $\downarrow$ & Rate of novel words & Repetitive language indicates exhaustion \cite{Harrison97}.\\
\midrule
\multicolumn{5}{@{}c}{\textbf{\textit{Disengagement} — withdrawal and distancing from work \cite{demerouti2001job}}} \\
\cmidrule(l{0pt}r{0pt}){1-5}
\addlinespace[2pt]
Commit Volume (CV)       & Commits & $\downarrow$ & Monthly commit count & Declining output reflects withdrawal \cite{Afrahi22}.\\
\addlinespace[2pt]
Review Participation (RP)& Reviews & $\downarrow$ & Monthly review count & As above, for review activity~\cite{Afrahi22}.\\
\addlinespace[2pt]
PR Rejection Rate (RR)   & PRs & $\uparrow$ & Ratio of PRs closed without merge & Rising rejection reflects declining competence, indicating disengagement \cite{Austin19}.\\
\addlinespace[2pt]
Task Abandonment (TA)    & PRs & $\uparrow$ & Proportion of PRs opened and never closed & Direct signature of disengagement \cite{Miller19}.\\
\addlinespace[2pt]
Pronoun Use (PU)         & Texts & $\uparrow$ & Log ratio of singular to plural first-person pronouns & Shift toward singular pronouns indicates distancing from team \cite{Lyons18}.\\
\addlinespace[2pt]
Lexical Disengagement (LDs)   & Texts & $\uparrow$ & Mean similarity to disengagement seed phrases (\S\ref{sec:text})& Explicit detachment language reflects disengagement. \\
\addlinespace[2pt]
Sentiment--Disengagement (SD)  & Texts & $\uparrow$ & Negative sentiment score (fine-tuned RoBERTa, \S\ref{sec:text}) & Disengagement associates with increased negative sentiment \cite{raman2020stress}.\\
\bottomrule
\end{tabular}
\end{table*}




\subsection{Text-derived Signals}
\label{sec:text}
Two of the studied signals required more than lexical/timestamp analysis. In such cases, we used learned models to capture nuances in contributor texts.

\textbf{Sentiment}: General-purpose sentiment classifiers are unreliable on software engineering text. Domain terms such as ``kill'' and ``execute'' are neutral in technical usage but are scored as negative by classifiers trained on general English. Even classifiers trained on labeled software data have reported F1 scores between 0.15 and 0.64 on the positive and negative classes~\cite{Biswas20}. Following the methodology of Biswas et al.~\cite{Biswas20}, we fine-tuned a RoBERTa model (instead of the original study's BERT model) for 3-class sentiment classification (positive/negative/neutral). This fine-tuned model exceeded the original BERT model's macro-F1 score when using the same dataset proposed by Biswas et al., achieving 0.889 across 10-fold cross-validation compared to the original's 0.870. 

\mariam{to save space this paragraph can be shortened}
\tim{paragraph shortened}
\textbf{Lexical Features}: To capture exhaustion and disengagement language, we follow a similarity-based unsupervised text classification method that embeds text and category descriptions into a shared space and compares via cosine similarity~\cite{Wang2022}. We created 30 seed phrases for each dimension (e.g., ``completely drained'', ``running on empty'' for exhaustion, and ``not going to review'', ``you decide'' for disengagement) and encoded both sets with a pretrained MiniLM~\cite{Wang2020} sentence transformer. Each contributor message is split into sentences. For each sentence, we take the mean cosine similarity to its three closest seeds in a category, and average across sentences to obtain the message's score for that category. For a sanity check, we manually reviewed 90 scored messages sampled evenly across the similarity distribution and judged each as a genuine match, a plausible-but-ambiguous match, or a false positive. 67.78\% of high-similarity messages were judged genuine matches, with sarcasm being the leading source of false positives.

\subsection{Per-signal Scoring}
The framework begins by scoring each signal individually. We orient each raw metric to map higher values to greater risk of burnout. 
Signals marked $\uparrow$ in Table~\ref{tab:catalogue} are used as-is, and signals marked
$\downarrow$ are inverted. 
Among all signals, Sentiment-Exhaustion captures flattened sentiment, rather than simply high or low sentiment. Since exhaustion can manifest as a decrease in emotional expression~\cite{Porr2010}, what matters is not whether a message reads positive or negative, but how intense that sentiment is. We therefore take the absolute value of each sentiment score, discarding its sign and keeping only its magnitude. We then invert the resulting score, so that flatter sentiment (lower magnitude) indicates higher risk.

We then compute two separate components: a level score (percentile ranking of the current value against all prior values for a contributor within a given signal) and a trend score (the percentile rank of a rolling 4-month linear regression slope). 
Combining the level score and trend scores relative to a contributor's own history, this parallels methods used in behavioral monitoring for mental health, where risk is estimated by looking at a person's deviation from their own established pattern \cite{Fang2026}. The two components for each score are averaged into a single score per signal, then smoothed over a 4-month exponentially weighted moving average to reduce noise from any single month. This value is then passed through a sigmoid function (center=0.65, steepness = 8). The center is offset as under a standard sigmoid centered at 0.5, a contributor performing exactly as their history would predict (raw score of 0.5) would map to an elevated output; moving the center to 0.65 means a contributor must be performing above their own historical norm to score as elevated. The steepness is increased to sharpen contrast between elevated and non-elevated contributors. This pipeline produces a relative, monthly score for each contributor, and is repeated across a contributor's entire lifetime. %

\subsection{Dimension Aggregation and BRS}
Scored signals are combined within each OLBI dimension as a weighted average, with weights normalized to sum to one within the dimension:

\begin{equation}
E = \sum_i w_i^E \cdot s_i \qquad D = \sum_j w_j^D \cdot s_j
\label{eq:aggregation}
\end{equation}

Where $i$ is each proxy variable for exhaustion, and $j$ is each proxy variable for disengagement. $w^E$ and $w^D$ refer to the weight vector for each of the dimensions, respectively. $s$  refers to the value of a specific variable at a given month. \mariam{perhaps you meant to say "refers to the value of a specific variable at a given month"}
\tim{I agree, sounds better}The monthly BRS is the mean of the two dimensions, 
$BRS=(E+D)/2$. 

The equal weighting across dimensions is a commitment to OLBI's structure, which treats exhaustion and disengagement as equal components of the syndrome rather than weighting one above the other.

Within each dimension, weights are learned rather than assigned. Signals differ in how directly they reflect their dimension: commit volume is a near-direct signature of disengagement, while rejected pull requests are a more distant one. Rather than encode our own intuitions about this ordering, we fit the weights against labeled cases. Section~\ref{sec:fitting} describes the fitting procedure and the sensitivity of the resulting weights to initialization.


\section{Preliminary Evaluation of BurnRiSc}\label{sec:study_design}
\subsection{Research Questions}
We conduct a preliminary evaluation of BurnRiSc against three questions.

\noindent\textbf{RQ1 (Validity). Does BRS elevation precede a contributor's public disclosure of burnout, and by how long?}
BurnRiSc's premise is that OLBI's dimensions leave traces in repository activity prior to a contributor's self-disclosure of burnout. 
We therefore ask whether BRS crosses the risk threshold (\S \ref{sec:riskonset}) in the months before a public disclosure, and how far in advance. Lead time matters as much as coverage: an instrument that registers a case only once it is publicly evident restates what is already known, whereas one that registers it months earlier is actionable. 
We also report the cost side, the number of comparison contributors whose scores rise without any corresponding disclosure or activity collapse, since coverage without that figure is uninterpretable.


\mariam{Answer for Yuan, yes I think we should mention the validation against OLBI as a limitation and future work item.}
\tim{Added a portion in future work, let me know if I should quickly mention this further work here as well}
\vspace{0.1cm}
\noindent\textbf{RQ2 (Signal importance and stability). Which signals carry the most weight after fitting, and do the same signals dominate regardless of where the optimizer starts?}
The fourteen signals were derived from psychological and linguistic theory, but theory does not specify how much each should count, so the weights $w^E$ and $w^D$ in Eq.~\ref{eq:aggregation} are fitted rather than assigned (\S\ref{sec:fitting}). This raises an obvious concern: with ten confirmed cases and fourteen free parameters, a fitted solution could reflect our starting assumptions rather than the data. We therefore fit using four distinct normalization anchors and ask whether they converge. Similar weights from unrelated anchor points would indicate the weights are a property of the labeled cases; highly varying weights would indicate the fit is under-determined and the resulting weights should not be interpreted. Beyond stability, the fitted ordering is substantively interesting in its own right, since it indicates which observable behaviors carry the most information about exhaustion and disengagement, and thus which signals a future, leaner instrument would need to retain.

\vspace{0.1cm} 
\noindent\textbf{RQ3 (Discriminant validity). Is BRS separable from a contributor's contribution profile?} Several BRS signals require text and activity to compute. Indeed, contributors with a richer contribution profile generate more raw data for the framework. This creates a specific failure mode: the score could rank prolific contributors as at risk simply because they are prolific, reproducing a measure of activity under a well-being label. Such an instrument would direct attention toward a project's most productive members regardless of their actual state. We therefore correlate peak BRS against an \textit{activity score} (\S\ref{sec:activity}) constructed to share no inputs with any of the fourteen signals. The expected relationship is not zero, as sustained high output in an unpaid setting is itself a plausible antecedent of exhaustion, so the construct predicts some association. However, this association should be weak enough that BRS is not recoverable from activity alone. We additionally identify contributors whose scores are elevated despite sparse activity, where an elevated score rests on too little evidence to be trusted.

\mariam{Response to Yuan: it would be good to provide statistics on the project in supplemental. We don't have to name the project but we can use maturity levels, domain, popularity, age, contributor count, etc... }\yuan{I agree, tim will add that to the supplementary document}


\subsection{Subject Selection} \label{sec:subject_selection}
We selected ten GitHub repositories, each containing at least one contributor who publicly disclosed burnout (\S\ref{sec:confirmed_case}). A supplementary document in our replication package reports per-repository statistics, with repositories identified by number. We withhold names and handles to avoid re-identifying the disclosers.

We included repositories where at least one contributor self-reported burnout. We chose this filter to enable within-repository comparison between contributors who self-reported burnout and those who did not. The rarity of self-reported cases of burnout on GitHub resulted in a limited dataset for our preliminary analysis. From each repository, we chose the ten most prolific contributors by commit count, together with contributor who self disclose burnout and fell outside that set. This yielded \mariam{"100 candidates" only holds if every discloser was already in their repo's top 10, can we confirm? If the case we can state it explicitely} \tim{Each confirmed case was in the top 10, added explicit statement} 100 candidates, as each confirmed case fell in the top ten all-time contributors of their repository. Each repository has a different size, and thus there is a large variety of how many commits the average contributor within a given repository makes. As such, the top ten are chosen so we have a variety of contributors from a variety of repositories. Using a commit cutoff threshold may cause larger repositories to overfill the dataset while leaving smaller repositories with only 1 or 2 contributors. For each selected contributor, we extracted five categories of timestamped activity through the GitHub API: commits with message text, pull request creation and merge events, code reviews, pull request review comments, and issue comments. We then applied the two applicability conditions specified in Section~\ref{sec:framework_condition}.  The resulting study set comprises 68 contributors across ten repositories. 

\subsubsection{Confirmed Cases of Burnout}\label{sec:confirmed_case}
A contributor was associated with a confirmed case of burnout when two criteria were met: an explicit public statement of burnout (found through GitHub API keyword search such as tired, burntout, burnout, leaving) and manually verified) and at least one year of recorded activity \mariam{question for Tim: do you mean before the statement here? I also suggest changing statement -> self disclosure of burnout} 
\tim{yes, adjusted to self-disclosure.}before self-disclosure. We dated a contributor's burnout to the month containing the disclosure post. This is a conservative choice, as burnout may have occurred before self-disclosure. Ten contributors met these criteria.

\subsubsection{Activity-Collapse Cases}
Ten confirmed cases are too few to fit fourteen weights, so we constructed a second, weakly labeled group: contributors whose contribution volume collapsed in a pattern resembling the confirmed cases. We do not treat these as burnout cases; they serve as an intermediate group between the confirmed cases and the remaining contributors, allowing the fitting procedure in Section~\ref{sec:fitting} to target an ordering rather than a binary split.

\paragraph{Identifying activity collapse.}
For each selected contributor that did not have a confirmed case of burnout (58 contributors), we computed monthly activity and anchored a baseline to their historical peak, defined as the maximum of the trailing-window median over their full history. A conventional trailing baseline is unsuitable here: when a contributor declines gradually, the trailing average declines with them and no drop is ever large enough to flag. Anchoring to the peak mirrors the drawdown measure used in financial risk management, where decline is assessed against the highest point reached rather than a recent average~\cite{Mahmoud16}. We flag a drop when monthly activity falls to 50\% or less of this peak for at least two consecutive months, a simplification of changepoint detection~\cite{Killick12}, which has previously been applied to OSS activity series~\cite{Walden21}.

Severity of a flagged drop is assessed over a fixed six-month post-onset window, following the logic of interrupted time series analysis, which compares bounded pre- and post-event periods rather than a lifetime average that could conceal initial severity behind a later partial recovery. For each candidate:
\[\mathit{severity} = \mathit{dropPCT} \times \mathit{durationFactor} \times \mathit{baselineFactor}\]

$dropPCT =1 - (postDrop-onsetAverage) / baseline)$, which aims to capture how deep the drop is.

$durationFactor=min(duration,12)/12 $ captures persistence, with duration measured as the months elapsed before activity recovers above 70\% of baseline and capped at one year so that indefinite inactivity does not dominate the ranking.

$baselineFactor=log(1+baseline)$ favors contributors who were consistently active beforehand over those who were marginally active and went quiet.

We retained only the most severe drop per contributor. Twelve contributors were selected as part of the activity collapse group, leaving 46 comparison contributors with neither a disclosure nor a qualifying collapse. 

\subsection{Weight Fitting}
\label{sec:fitting}
Within-dimension weights are fit against the three labeled groups: confirmed cases, activity collapse cases, and comparison cases (46 contributors with neither a disclosure nor a qualifying collapse). For a candidate weight vector, we compute each contributor's lifetime peak BRS, since confirmed and collapse cases exhibit peaks near their disclosure or drop, respectively. The objective maximizes the combined separation between adjacent group medians in the ordering comparison → activity collapse → confirmed. Medians rather than means limit the influence of individual outliers, and targeting an ordering rather than a binary split uses the intermediate group as evidence about the direction of the score.

With ten confirmed cases, unconstrained fitting overfits. We add an L2 penalty pulling weights toward an anchor (regularization strength = 0.1), generalizing ridge-style shrinkage~\cite{Hoerl1970}. Each case starts with equal weights, but is pulled towards a different regularization anchor. We evaluate four anchors. \emph{Equal} assigns unit weight throughout. \emph{Heuristic} encodes how directly each signal reflects its dimension (higher anchor weights for off-hours activity, commit volume, lexical features, and task abandonment, lower weights for sentiment, lexical diversity, hapax rate, and pronoun use). \emph{Disengagement-heavy} and \emph{exhaustion-heavy} scale the heuristic prior's disengagement or exhaustion weights by two, respectively. Comparing across anchors tests whether the fitted solution is a property of the data or of our starting assumptions.

Regularization strength follows a one-standard-error style rule: we select the strongest penalty still achieving 90\% of the best observed separation~\cite{Hastie2009}. The penalized negative margin is minimized by L-BFGS-B~\cite{Byrd1995}, with each weight bounded to [0.05, 5.0] prior to within-dimension normalization, ensuring every signal retains some influence without any single signal dominating.

We validate the fit in two ways. Leave-one-out cross-validation over confirmed cases refits weights with each confirmed contributor held out. One hundred bootstrap resamples refit on resampled data to identify which weights are stable and which move with the sample~\cite{Hastie2009}.

The performance of the optimization was evaluated through the separation margin of the peak weights, defined as:
\begin{equation}
M(w) =    \big(\text{med}(R_w)- \text{med}(C_w)\big)+ \big(\text{med}(B_w) - \text{med}(R_w)\big)
\label{eq:seperation}
\end{equation}
where $C_w$, $R_w$, and $B_w$ are the peak BRS distributions of the comparison, activity-collapse, and confirmed groups under weight vector $w$, and $med()$ denotes the median. $M(w)$ is exactly the quantity the fitting procedure maximizes.


\subsection{Risk Onset and Peak Criteria}
\label{sec:riskonset}

RQ1 asks whether BRS elevation precedes disclosure and by how long. Answering it requires comparing two dates, but BRS supplies none: the framework specified in Section~\ref{sec:framework} assigns a BRS score to every month and never declares that a contributor has been flagged. We therefore define two ways of reducing a contributor's monthly series to a single dated event, each capturing a different sense in which the score may have registered an episode. Neither is part of BurnRiSc itself; both are decisions about how to read its output, and we report them separately so that the coverage attributable to each remains visible.

The first follows from the construct. Burnout does not occur suddenly; it builds up over time and comes with various behavioral changes. Thus single-month BRS spikes are not signals of burnout, while prolonged raised BRS levels are a possible indicator of burnout. We accordingly define \textit{risk onset} as the earliest month in which BRS exceeds the threshold and remains above it for at least two further consecutive months, dated to the first month of that run. \tim{added reasoning behind the threshold} \mariam{thx Tim, reads good to me} The threshold is derived from the fitted weights' peak distributions as the midpoint between the comparison group's 75th percentile and the activity collapse group's 25th percentile. This is centered between the upper edge of typical, non-burnout peak scores and the lower edge of activity collapse peak scores. The median for the activity collapse group was lower than that of the confirmed burnout group, thus we use this threshold as a way to ensure confirmed cases are caught. Anchoring to these tails rather than to a given median places the cutoff inside the region where the two groups' peak scores overlap, so it is not pulled toward whichever group has a more extreme median. As more data is added, both the weights and threshold should be re-evaluated to generalize better at larger scales. The persistence requirement filters single-month noise: an unusually heavy or unusually quiet month that crosses the threshold alone does not constitute a behavioral shift. At most one onset is recorded per contributor per year, taking the earliest qualifying run, since OSS contribution is intermittent enough that an unconstrained detector fires repeatedly on what is plausibly a single underlying episode.

Persistence buys specificity at a cost. An episode that is severe but short, or that resolves before three months accumulate, produces no onset even where the score rose sharply. We therefore admit the \textit{contributor's lifetime peak BRS} as a secondary criterion, which is insensitive to duration by construction and so recovers exactly the cases persistence discards. The peak also serves as the per-contributor summary statistic targeted by the weight fitting (Section~\ref{sec:fitting}) and correlated against the contribution profile in RQ3.
\subsection{Activity Confound Testing} 
\label{sec:activity}
To answer RQ3, we compute an activity score for each contributor from event counts alone, sharing no input with the fourteen BRS signals. The activity score serves two purposes: (1) It checks the quality of the data of a given contributor, that is, are they contributing more than just commits, how consistently are they contributing, etc. ; (2) It normalizes event volume to a more manageable scale, as raw event scores are heavily right-skewed and are dominated by outliers. As such, the activity score paints a clearer picture of how involved a contributor is compared to pure activity volume.

The activity score combines volume (log-scaled monthly event mean, so that outliers do not dominate the scale), tenure (fraction of months with any activity), consistency (one minus the coefficient of variation, favoring sustained contribution over bursts), and diversity (fraction of the five tracked activity types the contributor engages in at all), weighted 0.4, 0.25, 0.2, and 0.15, respectively, producing a score in [0,1].

\section{Results}\label{sec:result}
\subsection{\textbf{RQ1:} \textbf{Validity}}
To evaluate burnout prediction, we built each contributor's monthly BRS series using the pipeline described in Section~\ref{sec:framework}. Once we construct a contributor's full series, we scan it for risk onset and the BRS peak. For confirmed cases, we ask whether an onset falls within the 24 months preceding disclosure, and record the lead time. 
We adopt a 24-month lookback window, noting that lookback period selection has no universal standard and should be justified transparently for the outcome at hand~\cite{Chisholm2026}.

We additionally record whether lifetime peak BRS exceeds the threshold within that window, since a peak may register an episode that never sustains three consecutive months. For collapse cases, we apply the same criteria relative to the most severe drop. For the 46 comparison contributors, we record how many exhibit an onset or a threshold-crossing peak at any point, which gives the cost side of the detection claim.

For the confirmed cases (Table~\ref{tab:signals}), BurnRiSc flags 6 of 10 contributors who self-report within 6--15 months. Admitting the lifetime peak month as a secondary criterion raises coverage to 8 of 10, at lead times of 5–16 months. The two contributors falling outside the two-year window, Dev 5 and Dev 8, are not missed outright: both are flagged prior to disclosure, and Dev 8 produces a fresh flag upon returning to the project. The underlying signals therefore appear directionally correct for all ten cases, even where the timing falls outside the window we treat as actionable.

Of the 12 activity collapse cases, 6 were flagged within 4 to 13 months before their greatest activity decline, rising to 10 out of 12 when using peak BRS as a secondary criterion. The increase in identification rate is similar to what occurs with the confirmed cases, indicating that the two-criterion approach generalizes outside of the confirmed burnout set. Since the peak signal tends to fall close to the time of self-reported burnout or activity collapse, it is used as a heuristic for optimization and confound analysis in Sections~\ref{sec:rq2_result} and \ref{sec:rq3_result}, respectively.

Among the 46 comparison contributors, 35 (76.08\%) never cross the threshold. The remaining 11 cannot be treated as errors outright. These contributors neither disclosed burnout nor exhibited a qualifying activity collapse, but the absence of both is not evidence that they were unaffected. We regard roughly one crossing in five as reasonable, given that OSS contribution patterns are highly irregular and that the persistence and once-per-year constraints are tuned to tolerate some noise rather than to flag every BRS spike.

Taken together, BurnRiSc's coverage is 8 of 10 on confirmed cases and 10 of 12 on collapse cases when peak BRS is used as a secondary signal. These results indicate BurnRiSc performs fairly well as an early-warning framework for OSS burnout. It reliably flags at-risk contributors with an actionable lead time in a large majority of cases, while keeping potential false alarms on comparison contributors within an acceptable range.

\begin{table*}
\setlength{\abovecaptionskip}{4pt}
\caption{Confirmed Burnt-out contributors. Each contributor is accompanied by when they were flagged by the framework, the month they scored their highest BRS, and the number of months between the flag/peak and the disclosure date. The Flag to Burnout column uses the flag closest to the disclosure date that still precedes it.}
\label{tab:signals}
\small
\begin{tabular}{|@{}>{\centering\arraybackslash}p{0.1\linewidth}|>{\centering\arraybackslash}p{0.1\linewidth}|c|>{\centering\arraybackslash}p{0.1\linewidth}|>{\centering\arraybackslash}p{0.1\linewidth}@{}|>{\centering\arraybackslash}p{0.1\linewidth}|}
\hline
\textbf{Dev ID}& \textbf{Self-Report Burnout Date}& \textbf{Dates of Flags}& \textbf{Peak BRS Month}& \textbf{Flag to Burnout}&\textbf{Peak to Burnout}\\\hline
1& Dec 2016& Mar 2015, June 2016& June 2016& 6 Months&6 Months\\\hline
2& Mar 2017& Feb 2016& Oct 2016& 13 Months&5 Months\\\hline
3& Jun 2017& N/A& Jan 2017& N/A&5 Months
\\\hline
4& Feb 2020& Sep 2015, Feb 2019, Jun 2021, Jun 2024& Jul 2021& 12 Months&N/A\\\hline
5& Sep 2020& May 2017& Dec 2017& 40 Months&33 Months\\\hline
6& May 2022& N/A& Jan 2021& N/A&16 Months\\\hline
7& Jun 2022& Jun 2011, Jun 2018, Mar 2021& Apr 2021& 15 Months&14 Months\\\hline
 8& Feb 2023& Sep 2016, Oct 2018, May 2026& Jun 2026& 52 Months&N/A\\\hline
 9& May 2023& Oct 2022& Nov 2022& 7 Months&6 Months\\\hline
 10& Dec 2025& Sept 2020, Jun 2025& Nov 2020& 6 Months&61 Months\\ \hline
\end{tabular}
\end{table*}

\subsection{RQ2: Signal Importance and Stability} \label{sec:rq2_result} 
Signal weights are fit numerically, aiming to separate three labeled groups: confirmed cases, activity collapse cases, and comparison. The optimization was performed as described in Section \ref{sec:fitting}. After optimization, the separation margin (Eq. \ref{eq:seperation}) improved substantially with all normalization anchors compared to the unweighted baseline (0.121): Equal 0.243, Exhaustion-heavy 0.241, Disengagement-heavy 0.224, and Heuristic 0.187. Since the Equal-anchored fit achieved the best separation, these weights are used for all evaluations in this paper. The resulting risk threshold, derived from the Equal-anchored fit's peak distribution, is 0.412, which is also used for all evaluations in this paper.

These weights are validated in two ways. Leave-one-out-cross-validation, refitting, with each confirmed contributor held out and comparing their peak BRS against the 80th percentile (reflecting a similar false-positive tolerance already accepted in RQ1) of the training set's comparison-group peak distribution, finds 7 of 10 confirmed cases still rank above this floor when held out. Bootstrap resampling (100 iterations, refitting on each resample) finds the separation margin stable across resamples (mean = 0.215, std = 0.008), indicating the fit is not an artifact of the specific contributors sampled.

More informative than any single anchor's margin is which signals rank highest and lowest \textit{consistently}. \emph{Off-hours Activity} is weighted the highest on average across anchors (2.668), followed by \emph{Commit Volume} (2.416) and \emph{Lexical Exhaustion Features} (1.324). The lowest weights come from \emph{Review Participation} (0.092) and \emph{Rejection Rate} (0.089). This suggests off-hours activity and commit volume are the framework's most load-bearing signals, while review participation and rejection rate contribute comparatively little under the current data. The full per-metric weight vector under each anchor, along with the corresponding peak-risk distributions, is reported in the supplementary document.

The weights produced in this procedure show which features to focus on for future iterations of the BurnRiSc framework. However, we are cautious about discarding them outright: their low weight may reflect limitations of the current dataset (e.g., low review volume, small numbers of rejected PRs) rather than a genuine absence of signal, and their relevance could differ in organizations with heavier code-review cultures. We therefore recommend that these weights be re-validated on larger datasets before permanently deprecating low-weight features and examining potential redundancy among the higher-weighted features (e.g., whether \emph{the Lexical Exhaustion Feature} partially captures variance already explained by \emph{Off-hours Activity}) to guide a more robust next-generation feature set. In addition to the weights, the threshold should also be re-evaluated on a larger dataset alongside the weights, as it is based on the distribution created by a given set of weights.

\subsection{RQ3: Discriminant Validity} \label{sec:rq3_result}


Peak BRS correlates weakly with activity score (r = 0.260, r² = 0.068, p = 0.032): a real, statistically detectable association, consistent with the construct's prediction, but one accounting for a low amount of the variance in peak risk. Figure~\ref{fig:RQ3} plots peak risk against activity score, producing four quadrants: genuine signal (high activity, high risk), healthy contributor (high activity, low risk), too sparse to assess (low activity, low risk), and over-analysis risk (low activity, high risk). Only four contributors fall into the over-analysis quadrant, an elevated score built on too little\yuan{relatively sparse? we already selected top-10 most committed contributors..unless the project is too small, i am not sure if we could say we know too little about these contributors}\tim{unfortunately there was a couple of projects on the smaller end, thus even though theyre in the to 10, they still have little informatio} behavioral history to trust. 16 out of the 27 contributors with a peak risk above the threshold turn out to be either confirmed cases or highly-ranked activity collapse cases, further confirming the validity of the framework. The remaining 11 could include real undetected cases, so 16/27 is a lower bound on precision. 
\tim{added Mariams suggestion as final sentence}

\begin{figure}[t]
\label{fig:RQ3}
    \centering \setlength{\abovecaptionskip}{4pt}
    \includegraphics[width=0.45\textwidth]{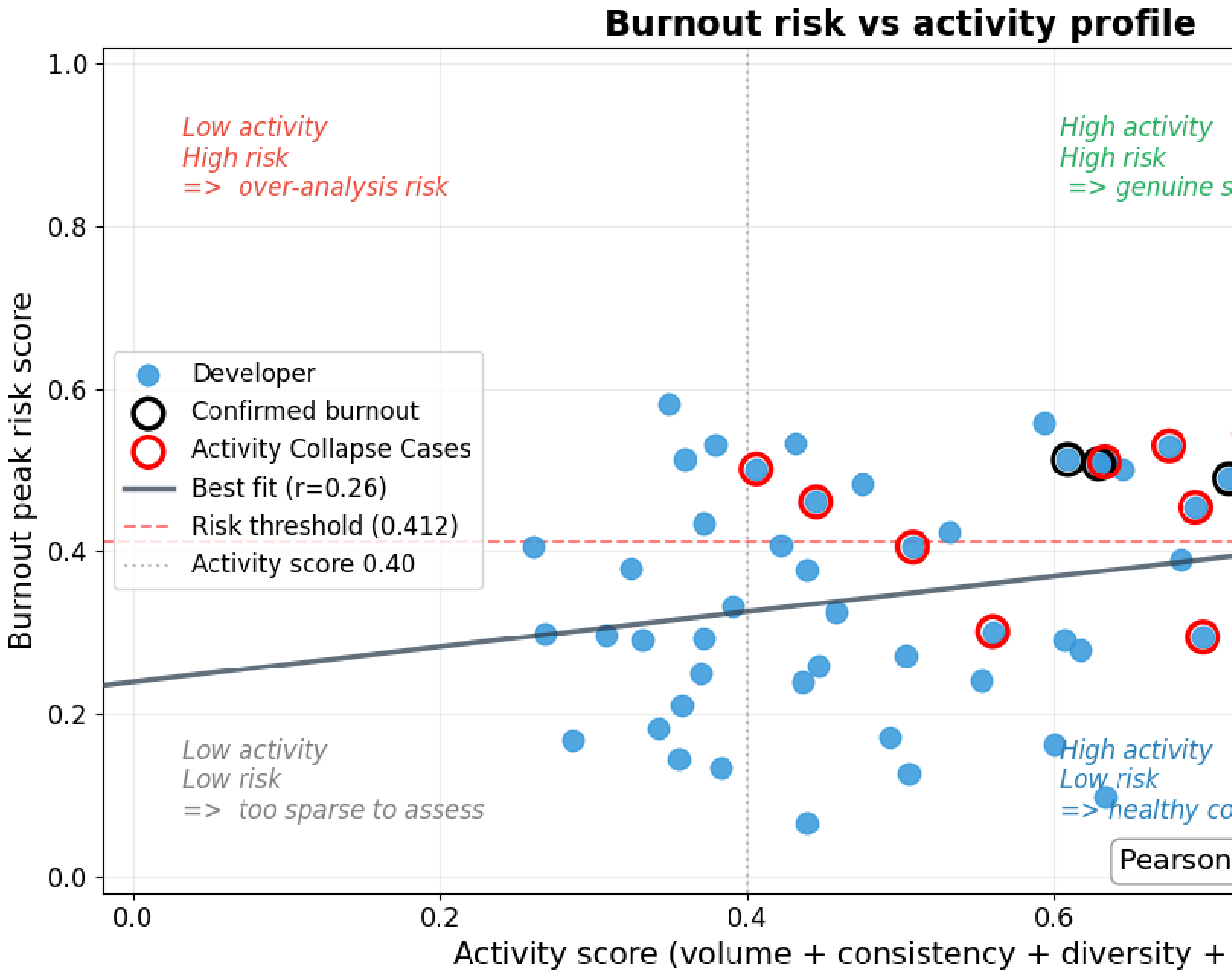}
    \caption{Peak BRS compared to activity scores. A line of best fit is added to show low correlation with the data.}
    \label{fig:RQ3}
\end{figure}

\section{Discussion and Further Work}\label{sec:disc_future}
\noindent \textbf{What the BRS is and isn’t. }BurnRiSc is a screening tool, not a diagnostic tool. The distinction matters as diagnosis requires using the OLBI itself, administered directly to a person, or through clinical judgment. BurnRiSc aims to produce a proxy for the OLBI by inferring public behavioral and linguistic traces that correlate to the dimensions of the OLBI. A high BurnRiSc score should prompt concern, not a conclusion. Treating it as anything greater is an overstatement of what the proxy variables can support.

\vspace{0.1cm}
\noindent \textbf{False positives are safer than false negatives.} 
\mariam{Our threshold (4.4) is the exact midpoint between the two group percentiles, neutral 50/50 split, not skewed toward catching more true positives. We should soften this claim or make clear what criteria justifies this claim} This framework's design choices trade more false positives for fewer false negatives. This is intentional, as false positives cost a brief, low-stakes check-in with a contributor who turns out to be fine, while a false negative means burnout goes undetected, the outcome the framework exists to prevent. Because our comparison group reflects absence of disclosure, not confirmed absence of burnout, some false positives may be undisclosed true positives. Our false-positive count is therefore an upper bound, which makes the precision we report a lower bound on the framework's true precision.
This tilt comes from signal weighting and from the classification threshold itself. The threshold sits at the midpoint between comparison cases and activity collapse cases, instead of comparison and confirmed cases. This ensures more burnout cases are caught by the framework, but it also results in more false positives.
\mariam{should we talk about the precision being at a lower bound? We are getting at most X many false positives, so precision can only be higher than what is computed. }\mariam{I believe this part still needs some wordsmithing}This has a direct deployment implication: any real-world use should pair a flag with a cheap, non-judgmental response, not a formal report or intervention. A flag should prompt a contributor to reflect how they are feeling, and if warranted, take a break. This framework should not trigger an evaluation initiated by someone else. 
\tim{added clarity on False positives, and threshold, rewrote last paragraph, and added discussion about lower bound}

\mariam{reworded a bit the implication paragraph}
\vspace{0.1cm}
\noindent \textbf{Implications for OSS projects and maintainers.} For open-source projects and contributors, the appeal of a framework such as BurnRiSc comes from not needing any surveys or one-on-one interactions with contributors. Instead, it works in the background on existing data. This makes it possible for projects to monitor and remain aware of burnout risks across their contributor base and improve both contributors' quality of life and the overall quality of code in the projects. However, BurnRiSc is not meant to be used as an evaluation tool of contributors' engagement. Indeed, if contributors' behavior affects their standing, they may mask the very signals that BurnRiSc depends on, defeating its purpose and denying them the support it was designed to surface. 
The goal of the framework is to identify people who need support, not to penalize them for needing it.

\vspace{0.1cm}
\noindent \textbf{Threats to validity.} Our confirmed cases are those who burned out and disclosed it publicly, and they may differ systematically from those who experience burnout silently. The fitted weights thus apply to disclosed burnout, and their transfer to silent cases is untested; the activity-collapse group is a partial hedge, but it is weakly labeled. Keyword-based burnout self-disclosure identification narrows the channel further, missing indirect, off-platform, and non-English disclosures. The three-year tenure condition (\S\ref{sec:framework_condition}) and top-ten-by-commit selection (\S\ref{sec:subject_selection}) restrict the study to established core contributors, leaving burnout among newcomers and peripheral contributors outside what BurnRiSc can currently score.

\tim{added section about comparing to direct OLBI application}
\mariam{@Tim the added comparison to OLBI works nicely}

\vspace{0.1cm}
\noindent \textbf{Future Plans.} We have planned several extensions that follow from the limitations of this study. The most immediate is re-calibration: weights and threshold were fit on a low number of contributors and ten confirmed cases suffice to demonstrate feasibility but not to fit fourteen weights or support robust statistics; expanding the confirmed set is a critical next step. Detection itself should combine sustained onset with peak-risk verification, trading additional false positives for fewer false negatives in line with the framework's stated asymmetry. Beyond these, other questions remain open. Signals computed from contributor-authored text assume a human author. However, contributors may write commit messages, reviews, and issue comments with the help of language models. These signals would then measure generated text rather than the contributor's own linguistic behavior. The direction of bias is unclear: mediation may mask exhaustion by regularizing language, or delegation may itself signal depleted capacity. A further limitation is validity against the framework basis. BurnRiSc's signals are designed to approximate OLBI's dimensions, but this evaluation validates BRS against disclosure timing rather than against OLBI scores directly administered to the same individuals. Collecting OLBI responses to correlate with BRS is a natural next step. Agentic AI coding assistants raise another concern: a contributor can sustain commit volume while disengaging. We plan to examine validity under AI-mediated contribution. 
Finally, we plan to develop the ethical framing this paper sketches, because inferring psychological state from public traces raises questions of consent and appropriate use. 

\mariam{adding soem general comments here: 
- could we justify the threshold placement (midpoint of 75th/25th percentiles)?
- could we provide more details on the confirmed burnout cases (e.g., example keywords used)
- if space permits we can expand on implication for maintainers and we can have a threats to validity section (if appropriate for JAW, @Yuan what do you think?)
- Threat to validity: 
    - the sample size 
    - self disclosure as a biased channel: people who self disclose might differ from the population (those who silently experience burnout)
    - the 3 years tenure excludes newer contributors}    \tim{example keywords added to \ref{sec:confirmed_case}}

\bibliographystyle{ACM-Reference-Format}
\bibliography{_main}

\end{document}